\documentclass[fleqn,10pt]{wlscirep}
\usepackage[utf8]{inputenc}
\usepackage[T1]{fontenc}

\usepackage{multirow}
\usepackage{float}
\usepackage{enumitem}
\usepackage{amsmath,amssymb}

\title{Continuous Determination of Respiratory Rate in Hospitalized Patients using Machine Learning Applied to Electrocardiogram Telemetry}

\author[1,2, *]{Thomas Kite PhD}
\author[3]{Brian Ayers MD MBA}
\author[1,4]{Nicholas Houstis MD PhD}
\author[3]{Asishana A. Osho MD}
\author[3]{Thoralf M. Sundt MD}
\author[1,2,4,5,*]{Aaron D Aguirre MD PhD}
\affil[1]{Cardiology Division, Massachusetts General Hospital, Harvard Medical School; Boston, MA, U.S.A.}
\affil[2]{Wellman Center for Photomedicine, Massachusetts General Hospital; Boston, MA, U.S.A.}
\affil[3]{Cardiac Surgery Division, Massachusetts General Hospital, Harvard Medical School; Boston, MA, U.S.A.}
\affil[4]{Healthcare Transformation Lab, Massachusetts General Hospital; Boston, MA, U.S.A.}
\affil[5]{Center for Systems Biology, Massachusetts General Hospital; Boston, MA, U.S.A.}

\affil[*]{tomkite57@gmail.com, aaguirre1@mgh.harvard.edu}


\begin{abstract}
Respiration rate (RR) is an important vital sign for clinical monitoring of hospitalized patients, with changes in RR being strongly tied to changes in clinical status leading to adverse events. Human labels for RR, based on counting breaths, are known to be inaccurate and time consuming for medical staff. Automated monitoring of RR is in place for some patients, typically those in intensive care units (ICUs), but is absent for the majority of inpatients on standard medical wards who are still at risk for clinical deterioration. This work trains a neural network (NN) to label RR from electrocardiogram (ECG) telemetry waveforms, which like many biosignals, carry multiple signs of respiratory variation. The NN shows high accuracy on multiple validation sets (internal and external, same and different sources of RR labels), with mean absolute errors less than 1.78 breaths per minute (bpm) in the worst case. The clinical utility of such a technology is exemplified by performing a retrospective analysis of two patient cohorts that suffered adverse events including respiratory failure, showing that continuous RR monitoring could reveal dynamics that strongly tracked with intubation events. This work exemplifies the method of combining pre-existing telemetry monitoring systems and artificial intelligence (AI) to provide accurate, automated and scalable patient monitoring, all of which builds towards an AI-based hospital-wide early warning system (EWS).
\end{abstract}

\begin{document}

\flushbottom
\maketitle
\thispagestyle{empty}
\begin{figure}
\centering
\includegraphics[width=1.0\textwidth]{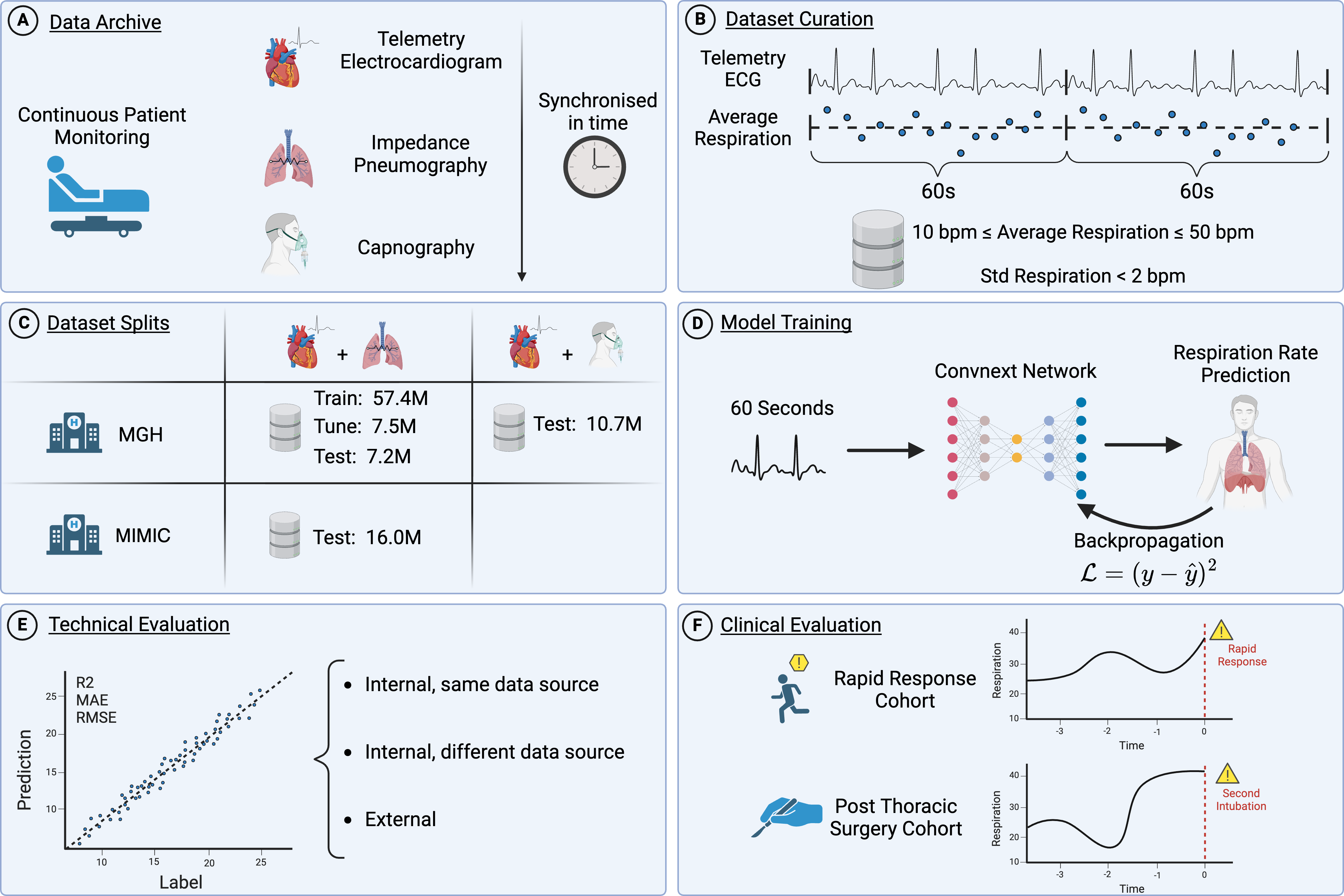}
\caption{Graphical abstract. A) A retrospective archive of standard vital signs and continuous telemetry waveforms is used for analysis. B) Downstream training and validation datasets are generated by aligning ECG telemetry waveforms with labels consisting of minute-long averages of automated respiration rate sources. C) Final datasets included a typical train/tune/test split from the same hospital and same respiration source, stratified by patient. Further validations include a distinct respiration source for ground truth respiratory rate from the same institution and a separate cohort from an external institution. D) A neural network is trained to predict the average respiration rate from the ECG telemetry segments, with training by backpropagating the mean square error. E) The neural network is \textit{technically} validated by inspecting labels and predictions for all test sets. F) The neural network is \textit{clinically} validated by studying two clinical cohorts: a cohort of patients that experience clinical deterioration requiring a rapid response event, and a cohort of patients who required a second intubation following thoracic surgery.}
\label{fig:graphical_abstract}
\end{figure}

\section{Introduction}
\label{sec:introduction}
The modern hospital setting is a highly collaborative and data rich environment which aims to closely monitor and promote the recovery of each and every hospitalized patient. Despite the sophistication of these environments, a persistently high number of patients suffer unexpected deterioration while under observation, leading to deaths that are potentially preventable\cite{rodwin:2020:preventable_mortality,makary:2016:medical_error}. It is tempting to tackle this problem through more intensive monitoring with a broader range of tests or with a higher frequency of assessments, but resource constraints already spread medical staff thin, increasing the possibility of error and raising the risk of burnout\cite{hall:2016:burnout_healthcare}. This work proposes an automated, inexpensive and scalable method for monitoring a key vital sign -- respiration rate\cite{cretikos:2007:rr_matters} (RR) -- on all hospitalized patients with access to standard electrocardiogram (ECG) telemetry. Such an approach would not only avoid burdening medical staff with new responsibilities, but could actually relieve them of manually recording RR, a measurement that is both time consuming and prone to inaccuracy \cite{lovett:2005:vexatious_vital, cretikos:2008:neglected_vital}.

Respiratory rate is a key determinant of blood oxygen delivery and carbon dioxide removal, one whose autoregulation can reveal important information about insults to a patient's respiratory and metabolic homeostasis. For example, a patient with hypoxic respiratory failure will breathe faster in an effort to defend blood oxygen levels. Similarly, a patient with a primary metabolic acidosis, perhaps from poor perfusion and rising lactic acid levels in the blood, will breathe faster to expel blood carbon dioxide and thereby compensate for this acidosis. Moreover, loss of respiratory control, as can result from neurologic injury or from medications that depress respiration such as opiates, can lead to inappropriately low RR and risk for hypoventilation. A single metric, RR, thus provides a window into a spectrum of physiologic processes, making it an essential vital sign to monitor patients for deterioration. For these reasons and more, RR is a commonly used predictor in hospital early warning systems (EWS) \cite{duckitt:2007:early_warning, moss:2017:manual_rr_for_ews}, defined here as collections of monitoring systems and algorithms to analyze patient data and predict deterioration as early as possible.

Devices for the automated monitoring of RR are available, but their use is often restricted to patients in the intensive care unit (ICU), limiting their applicability to a general hospital-wide EWS. Previous studies have demonstrated the potential to quantify respiratory rate from various biosignals (most frequently photoplethysmography, PPG, but also ECG) \cite{charlton:2018:breathing_review}.  In the hospital setting, ECG telemetry is a ubiquitous technology that allows for the continuous, noninvasive monitoring of each patient's cardiac electrical activity. Automated systems already use the ECG to track heart rate and rhythm, which allows for some simple alarms.  However, telemetry monitoring systems also capture the full high-resolution ECG waveform, beyond simple timing intervals, making it a promising target for the application of artificial intelligence (AI) algorithms to augment patient care\cite{kite:2024:telemetry_ecg}. Importantly for this work, the ECG telemetry signal contains variations attributable to patient respiration \cite{charlton:2017:rr_extraction}, which can be leveraged using a neural network (NN) trained to predict RR. Such an AI model could be connected to existing hospital telemetry systems, enabling real-time RR monitoring for almost all hospitalized patients.

In this work we train a model to predict RR from ECG telemetry by leveraging the two largest telemetry databases used in scientific work to date (to the best of the authors' knowledge). Technical and clinical validation establishes the model as both an accurate predictor of RR and a useful metric of clinical deterioration (Fig.~\ref{fig:graphical_abstract}). The results advocate for a future of AI-enhanced monitoring systems that could enable accurate and low-burden EWS's that improve patient outcomes.

\section{Dataset \& cohort characteristics}
\label{sec:dataset_and_cohort}
This work uses a retrospective archive of telemetry biosignals and vital signs collected from Massachusets General Hospital (MGH) between 2014 and 2023 (study approved under Institutional Review Board protocol 2020P003053). Respiratory rate measurements from either impedance pneumography (ImP) or capnography are time-aligned with single-lead ECG telemetry and divided into 60-second segments (a single lead is used as a single model input, taken from from any available physical leads, usually I, II, III and V). A comparable dataset was also curated from the open source MIMIC III waveform database \cite{Alistair:2016:mimic3_wfdb} as an external validation. More details of the selection criteria and curation are provided in Sect.~\ref{subsec:dataset_curation}. Key patient characteristics of the datasets and their partitions are given in Table~\ref{tab:cohort_stats}. The first three columns report patient-level information such as unique patient count, age distribution and patient sex. The last three columns report ECG segment-level information: total duration of telemetry signals, fraction of telemetry recorded in the ICU and fraction of telemetry recorded from intubated patients.
\subsection{Clinical cohorts}
\begin{table}[ht]
\centering
\begin{tabular}{lcccccc}
\toprule
Dataset & \# Patients & Age (mean $\pm$ std) & \% Male & Hours & ICU (\%) & Intubated (\%) \\
\midrule
MGH ImP Train         & 17441 & 62 $\pm$ 16 & 62\% & 239099 & 96\%  & 23\% \\
MGH ImP Tune          & 2219  & 63 $\pm$ 16 & 63\% & 31249  & 96\%  & 21\% \\
MGH ImP Validation    & 2182  & 63 $\pm$ 16 & 62\% & 29828  & 95\%  & 21\% \\
MGH Capnography       & 4832  & 64 $\pm$ 13 & 71\% & 45697  & 93\%  & 89\% \\
MIMIC                 & 4617  & 63 $\pm$ 16 & 56\% & 266702 & 100\% & -- \\
\bottomrule
\end{tabular}
\caption{Patient and segment level characteristics of the MGH datasets used for training and validation, as well as those of the MIMIC dataset used for external validation. ImP indicates impedance pneumography; ICU, intensive care unit.}
\label{tab:cohort_stats}
\end{table}

This work also studies two clinical cohorts to verify the practical utility of the trained model (Table~\ref{tab:patient_level_stats}). First, a rapid-response cohort, consisting of mostly standard medical ward (non-ICU) patients for whom an emergency response team activation was called, indicating the need for urgent medical attention. We filter this cohort to select for those patients undergoing respiratory failure (defined as requiring intubation within four hours of rapid response), and study their evolution. The second cohort involves a group of patients during the index hospitalization after undergoing cardiac surgery, available through the institutional Society for Thoracic Surgery (STS) database. A subset of the post-operative patients required a second intubation (reintubation) following the initial post-surgery extubation, most frequently indicating respiratory failure. A control group is assembled by selecting similar post-operative patients who did not require a second intubation. The control group is selected to match the time since surgery of the reintubated cohort in a 5:1 ratio (see panel A in Fig.~\ref{fig:sts_relative_change_composite})
\begin{table}[ht]
\centering
\begin{tabular}{lcccc}
\toprule
Cohort & \# Patients & Age (mean $\pm$ std) & \% Male & Event in ICU \\
\midrule
STS Test    & 66    & 65 $\pm$ 14    & 59\% & 97\% \\
STS Control & 316 & 64 $\pm$ 14 & 73\% & -- \\
\midrule
Rapid Response & 176 & 66 $\pm$ 16 & 64\% & 20\% \\
\bottomrule
\end{tabular}
\caption{Demographics of each clinical cohort and location of the event studied (ICU or standard medical floor). STS, Society for Thoracic Surgery.}
\label{tab:patient_level_stats}
\end{table}

\subsection{Respiration label distributions}
\begin{figure}
\centering
\includegraphics[width=1.0\textwidth]{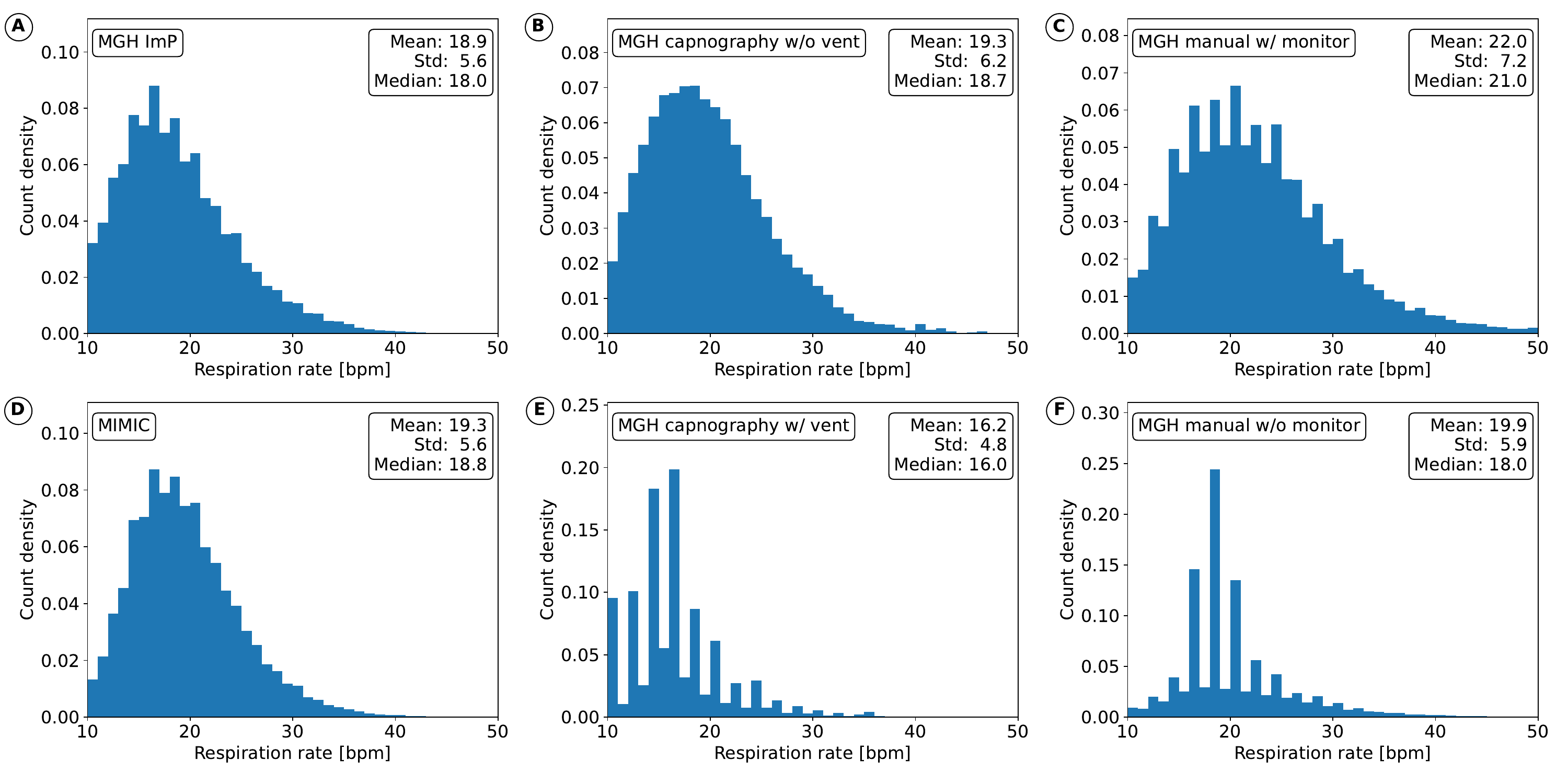}
\caption{Statistics of respiratory rate by data source. Histograms of respiratory rate measurements are plotted using integer bin sizes, facilitating comparisons to manual (human-based) measurements. The six panels reflect distinct combinations of dataset (MGH vs MIMIC), measurement modality (ImP, capnography, manual), and ventilator use (w/ or w/o).}
\label{fig:dataset_historgams}
\end{figure}

Fig.~\ref{fig:dataset_historgams} shows the distribution of respiratory rate values across the dataset and measurement sources used in this work. The model was trained using respiratory rate labels derived from the MGH ImP measurements (panel A). The MIMIC dataset (panel D) was used for external validation. The plotted capnography distributions were separated into those with and without active ventilator use (panel E vs panel B). The ventilator base rate is often set as an even integer, which explains the preponderance of even numbers in the distribution for ventilated patients. The manual, human-annotated, values are heavily biased by any available concurrent automated monitoring (e.g., ImP or capnography), in which case the number on the monitor is typically used instead (panel C vs panel F). When monitoring is not available the distribution is again heavily biased to even integers. Note that human-annotated values are not used within this work, but serve as an interesting demonstration of the applicability of the trained model (i.e. see also the manual RR annotations in panel C of Fig.~\ref{fig:full_technical_analysis}). One notable feature is that the MIMIC and MGH ImP data have subtly different shapes for low values, however the MGH capnography matches MIMIC statistics more closely. Training the NN on capnography may have yielded better generalization to MIMIC, however the ImP values were more plentiful (>5:1), which itself has large benefits for machine learning.

\section{Results}
\label{sec:results}

\subsection{Neural networks can accurately predict RR from ECG telemetry}
\label{subsec:technical_validation}
\begin{figure}
\centering
\includegraphics[width=1.0\textwidth]{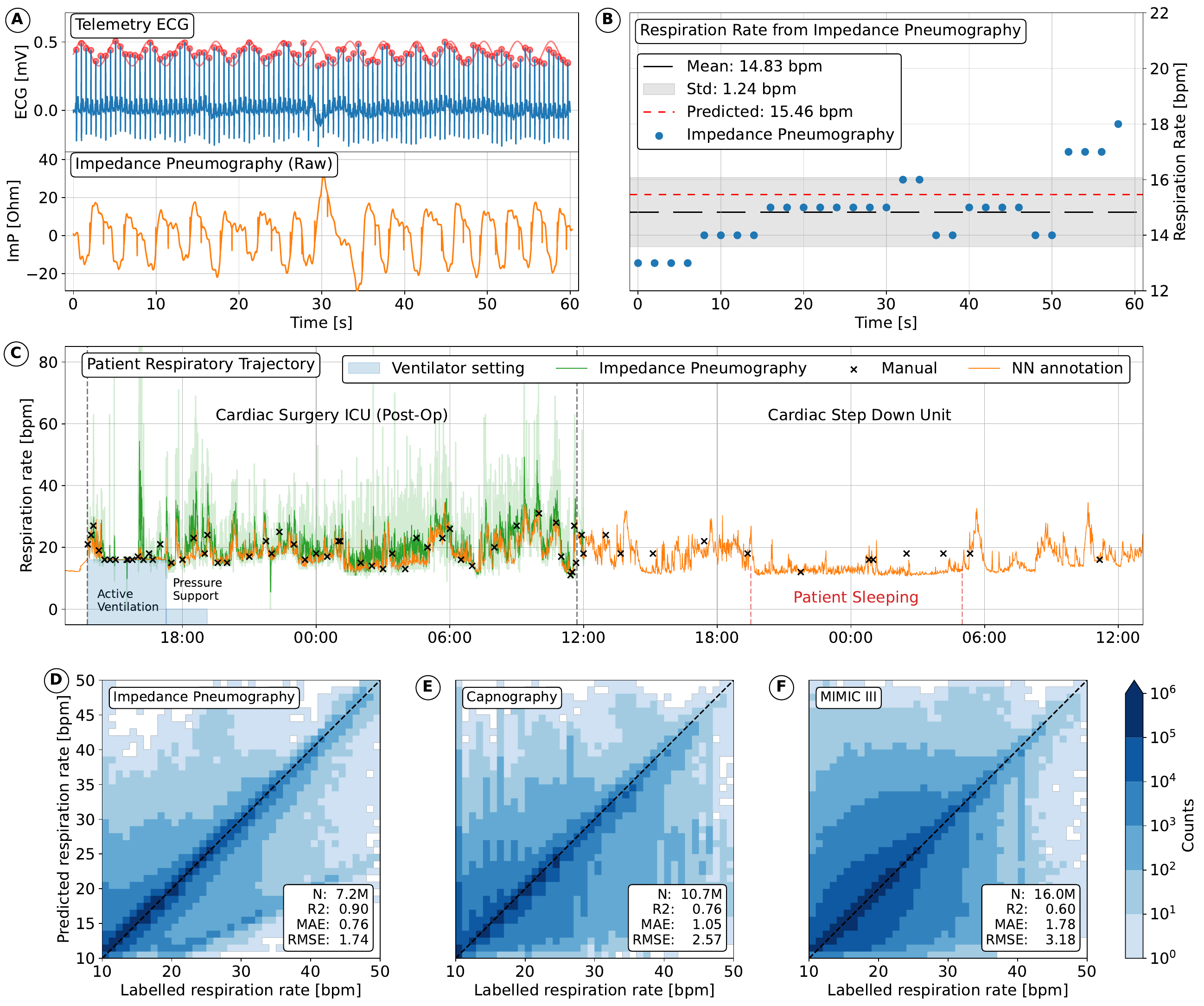}
\caption{Technical validation of neural network performance. A) An example of 60 second aligned ECG (blue) and impedance (ImP) respiratory rate (orange) waveforms with an illustrative breath-to-breath variation indicated with a sine wave (red). B) An example 60-second window with 30 respiration rate values associated with the ECG waveform arising from ImP. The mean and standard deviation are shown alongside the neural network prediction. C) An illustration of the neural network's ability to annotate respiratory rate longitudinally throughout patient's hospital course, notably in the absence of ImP once out of the ICU. D,E,F) Three evaluations of the neural network on three test sets: D) internal and same source labels (ImP) but new patients, E) internal but different data source labels (capnography), F) external but same source labels (ImP).}
\label{fig:full_technical_analysis}
\end{figure}

As described in Sect.~\ref{sec:dataset_and_cohort}, we leverage pre-existing automated sources of RR labels from ImP, and use only the most accurate of those to train a neural network (NN) to predict RR from ECG telemetry waveforms. RR labels from capnography are reserved to validate the network, alongside an external ImP based dataset. A total of almost 100 million minute-long paired ECG and RR measurements are assembled from two distinct hospitals. See Sect.~\ref{subsec:dataset_curation} for more information on dataset curation and filtering. Examples of input ECG, Imp labels, and model output are shown in Fig.~\ref{fig:full_technical_analysis} (panels A and B). In particular, Panels A and B show example waveforms (ECG and ImP) as well as the associated RR values (30 values sampled at 0.5\,Hz). An illustrative sine wave was overlaid on the input ECG (panel A) to qualitatively demonstrate ECG amplitude modulation during respiration, one of many known manifestations of breathing on ECG traces \cite{charlton:2017:rr_extraction}. Model architecture and training details are provided in appendix~\ref{app:nn_architecture}.

To illustrate the performance of the trained model in a monitoring context, we use it to annotate a patient's RR from their ECG telemetry spanning nearly 48 hours of care following cardiac surgery (Fig.~\ref{fig:full_technical_analysis}, panel C). As is typical for many institutions, immediately after the operation the patient received continuous ImP monitoring in the cardiac ICU, but when moved to a step down unit the next day, there was less monitoring (both automated and by medical staff). The patient was intubated and had a set RR on the ventilator for the initial approximately four hours in the ICU, before being transitioned to pressure support mode for weaning from the ventilator for an additional hour (blue regions). During that time and until the patient was transitioned out of the ICU 18 hours later, RR was available via ImP measurements (green lines). Rolling averages are shown alongside the data itself (solid vs transparent), demonstrating that the ImP measurements for RR are very noisy, even spiking beyond 100 breaths per minute (bpm). Human measured values (black ticks) show strong correlation with the ImP derived RR values while in the ICU. However, once transferred to a regular floor (non-ICU), the manually measured values become much more sparse (approximately every 4 hours) and are more uniformly around 18 bpm. Finally, the NN model's predicted RR (orange lines) can annotate the entire patient trajectory, revealing a much less labile RR compared to ImP measurements, and much more frequently than the manual assessments.

Panels D, E and F show the performance of the NN on multiple validation datasets: first, a held-out and blinded ImP-based internal MGH dataset split at patient level (i.e. patients in this validation dataset were not present in the datasets used for training and hyperparameter tuning). On this dataset the model's mean absolute error (MAE) is 0.76 bpm with a coefficient of determination (R2) of 0.90. Second, another internal MGH dataset based on capnography measurements, where the model's performance is largely preserved, though somewhat inferior to the validation ImP dataset, with a MAE of 1.05 bpm and R2 of 0.76.  Third, a completely external ImP-based dataset from the MIMIC III waveform database \cite{Alistair:2016:mimic3_wfdb}, arising from an independent hospital with different monitors. The model achieves an MAE of 1.78 bpm and R2 of 0.60 on this data, showing that the NN generalizes well to new patients, new sources of RR and even to new ECG monitors, albeit with more mismatch than the other comparisons. Notably the colors are logarithmically distributed, as the 1:1 line on linear display dominates the other bins by multiple orders of magnitude. The distribution of MIMIC data reveals that the two sets of ImP devices function quite differently. For example, the MGH devices have a tendency to capture a double frequency mode, oscillating twice as fast as the patient's true RR (confirmed via visual inspection for some cases). See Fig.~\ref{fig:dataset_historgams} for more information on label distributions.

\subsection{Continuous respiration rate measures patient deterioration}
\label{subsec:clinical_validation}
Given the demonstrated accuracy and generalizability of our RR estimation model, we next evaluate its potential utility within a clinical context. In particular, we aim to assess the model's potential to enhance patient monitoring and early warning systems through continuous tracking of respiratory dynamics prior to rapid-response and reintubation events.

\subsubsection{Rapid response cohort}
We first consider a cohort of patients for whom a rapid response event was initiated, triggered by patient decompensation that required immediate attention by the entire medical team. Specifically, we evaluate a group of 176 patients for whom the rapid response led to intubation within four hours of the event. All available ECG telemetry for these patients in the prior 37 hours is assembled and annotated with RR for each available 60-second segment (subject to filtering criteria discussed in Sec.~\ref{subsec:ecg_filter}). The nature of archived retrospective data means that it contains artifacts, noise and missing segments spanning both long and short time frames (i.e. missing some seconds or entire hours). As such, much of the analysis that follows gathers values into hour-long bins, and uses average respiration where >20 values were available (out of 60 possible minutes, 33\% data availability).

\begin{figure}
\centering
\includegraphics[width=1.0\textwidth]{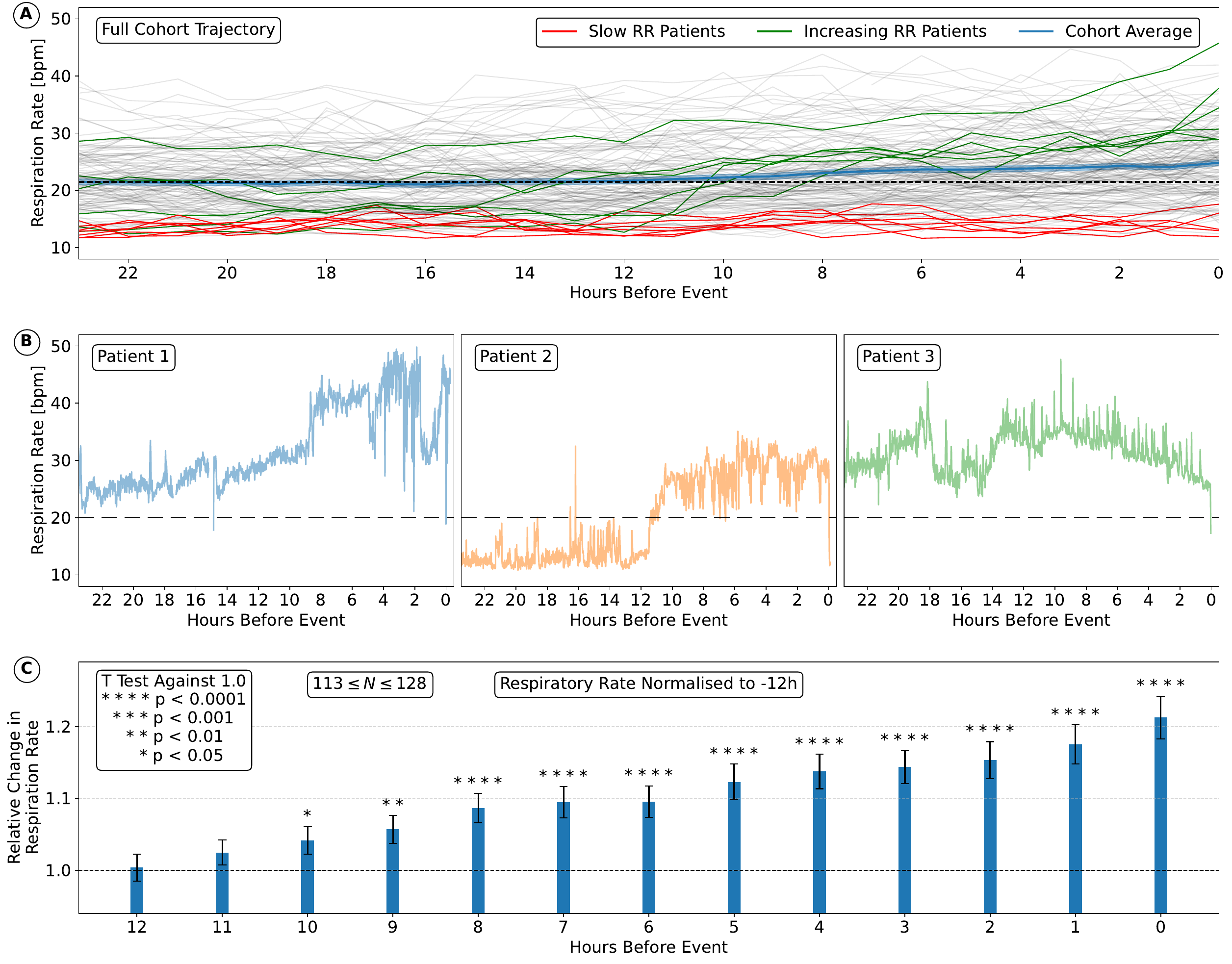}
\caption{Respiratory rate dynamics before rapid response events. A) A plot showing respiratory trajectories for the entire rapid response cohort, shown as hourly averages across time, with extreme behaviors shown in green/red and the mean in blue. B) Example minute-by-minute respiratory trajectories for three patients over the 24 hours leading up to a rapid response event. C) A statistical analysis of the change in RR prior to a rapid response event. For each hour leading up to the event a patient's average RR during the hour is compared to the average RR 12 hours prior, with the height of the bar reflecting the ratio. The number of patients who had usable telemetry signal for both hours varies by bar, with all bars having $113 \leq N \leq 128$. Asterisks indicate the statistical significance of a t-test comparing the observed ratio to a null hypothesis of $1.0$.}
\label{fig:floor_rr_relative_change_composite}
\end{figure}

Fig.~\ref{fig:floor_rr_relative_change_composite} shows the analysis for the rapid response cohort. Panel A shows the hourly average RR for each patient in the cohort leading up to the event. Some extreme behaviors are highlighted, demonstrating that many patients have a persistently low RR with no notable increase leading up to rapid response (red lines), while others have a dramatic increase in RR towards the event (green lines). These patients were selected for having all RR<18 bpm and a total RR increase of >14 bpm in 24 hours respectively. Despite the patients with persistently low RR, the cohort average (blue line, with shaded standard error) drifts significantly higher towards the time of decompensation, revealing a possible pathway towards early warning signs for those patients.

Individual RR trajectories, estimated minute-by-minute, provide additional insight into the relationship between respiratory rate dynamics and respiratory failure leading to intubation. Panel B shows three examples of such minute-by-minute trajectories, including gradual increase (patient 1), sudden step up (patient 2) and even gradual decline (patient 3) in respiratory rate. Thus there is no single, universal route to respiratory failure, but broadly speaking, changes in rate and increased variability are notable behaviors.

Panel C performs a statistical analysis of the data discussed above, identifying a strong relationship between change in RR and respiratory failure. Each patient's hourly RR in a given hour prior to the rapid response is normalized to their rate 12 hours before, thus providing a patient-specific baseline. This means a value of $1.0$ would indicate no change in RR, and consequently any cohort-wide difference from $1.0$ would be indicative of important physiological dynamics. The results show a statistically significant increase in RR starting 10 hours prior to the rapid response event, building to an average $\sim 20\%$ increase before the rapid response event. The statistical significances are t-tests against 1.0 (no change). 
In appendix~\ref{app:more_ref_hours} we explore more reference times beyond the 12 hours. The number of patients ($N$) can vary in each bar due to gaps in the available telemetry signal, but all bars had $113 \leq N \leq 128$ values. These observations underscore the clinical utility of continuous RR tracking within early warning systems, and highlight the potential of RR trends to enhance event detection compared to conventional fixed RR thresholds.

\subsubsection{Reintubation cohort}
\begin{figure}
\centering
\includegraphics[width=1.0\textwidth]{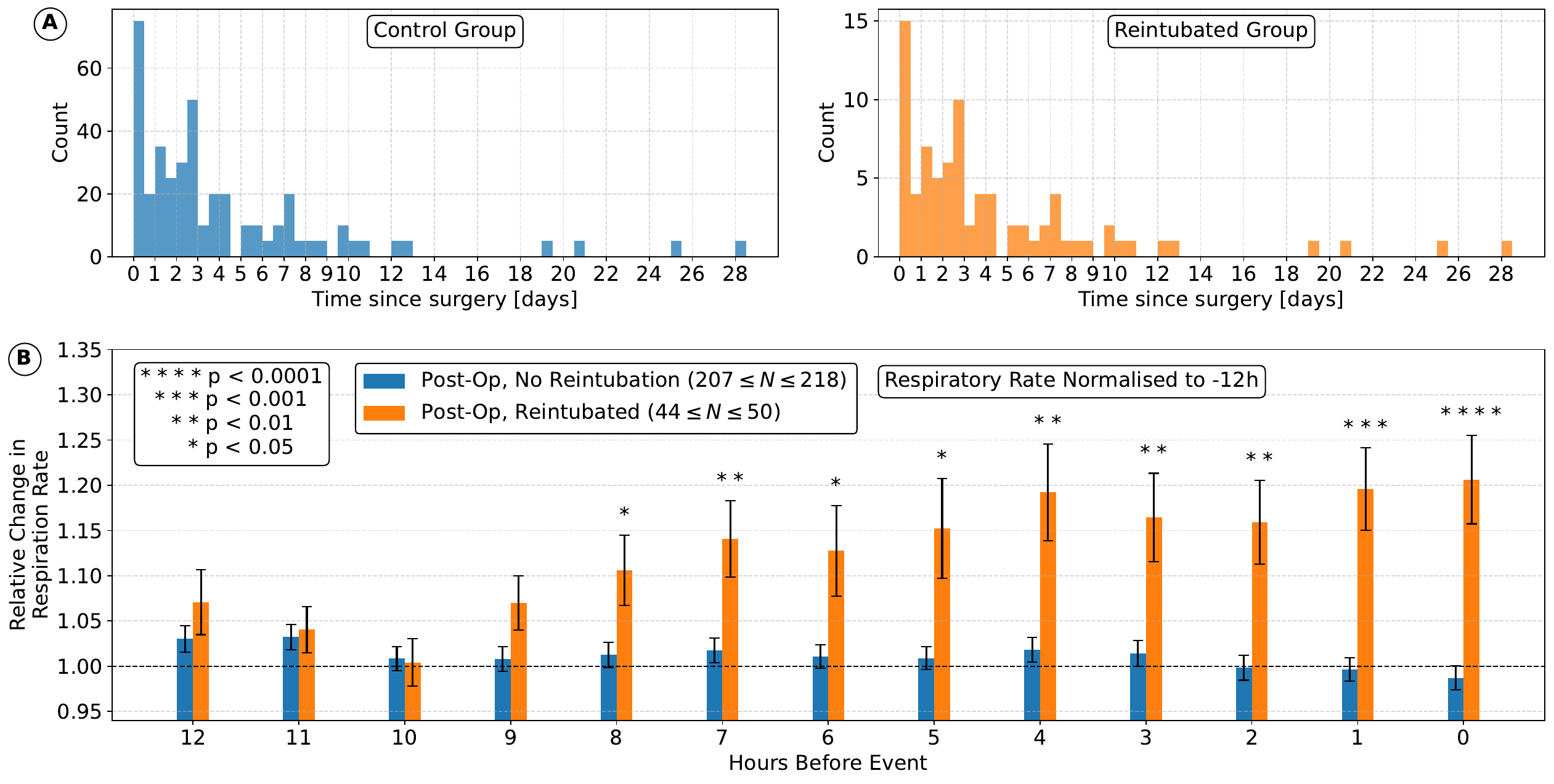}
\caption{Reintubation after cardiac surgery. A) Distribution of time elapsed between the end of surgery and reintubation (right) or time elapsed between the end of surgery and telemetry data collection in the control cohort (left). B) Statistical analysis showing the change in RR prior to reintubation, where the change is calculated as the ratio of a patient's RR averaged in a given hour prior to reinbutation (or matched time elapsed from surgery in the control cohort) divided by their RR 12 hours prior. The bars reflect the average ratio across the cohort. The number of patients, N, with usable telemetry during a given hour varied (range of N displayed). Asterisks indicate the statistical significance of a t-test comparing the two population ratios at each hour.}
\label{fig:sts_relative_change_composite}
\end{figure}

We assemble a second cohort of patients who experienced respiratory failure, drawing from the 5,571 patients in our institutional Society of Thoracic Surgery (STS) database, of whom 286 required reintubation in the post operative period during index hospitalization after cardiac surgery. To include only patients reintubated for respiratory failure (avoiding other indications such as need for additional procedures, etc.), patients were included in the reintubation cohort only if they had blood lab results indicative of respiratory failure (see Sec.~\ref{subsec:clinical_data_methods} for details). Of these patients, a total of 81 patients had sufficient telemetry leading up to the event. We also assemble a 5:1 control cohort of patients who did not require reintubation and who had telemetry recorded at matched times to the reintubation cohort (Fig.~\ref{fig:sts_relative_change_composite}, panel A). Matching the telemetry recordings by elapsed time since surgery enabled our analysis to control for RR trends attributable to typical postoperative recovery.

We next evaluate the relationship between individual RR trends and reintubation, similar to our approach with rapid response events. As before, we measure a patient's RR using the ECG telemetry model minute-by-minute, average the values in a given hour prior to reintubation, and normalize that value by a patient-specific baseline obtained from 12 hours prior. Reassuringly, the control group ratios are nearly flat across all lead times, as would be expected in the absence of a specific event towards which RR was evolving. In contrast, patients whose condition deteriorated sufficiently to require reintubation are found to begin breathing faster 8 hours prior to reintubation, again peaking at 20\% faster. (Fig.~\ref{fig:sts_relative_change_composite}, panel B). Thus, in a striking parallel to rapid response patients, postoperative patients displayed highly significant respiratory dynamics -- with similar physiological timescales -- for which continuous RR monitoring could enable more timely interventions.

\section{Discussion and Conclusion}
\label{sec:discussion_and_conclusion}
In this study we assembled a dataset of nearly 100 million minutes of ECG telemetry and used it to train and validate a neural network (NN) model capable of estimating RR, a key vital sign. This model was accurate to within 1 breath per minute (mean absolute error) on test data of ImP labels as well as capnography labels (a distinct measurement modality). It was accurate to within 2 breaths per minute on external data derived from a distinct institution and distinct ECG and ImP monitors. We subsequently used the model to continuously annotate RR across long tracts of retrospectively recorded ECG to study the relationship between respiratory dynamics and clinical events. In particular, we assessed the clinical utility of this model in two distinct patient cohorts: those experiencing emergency rapid response events (leading to intubation) on a hospital floor and those requiring reintubation following cardiac surgery. We observed a highly significant rise in predicted RR prior to respiratory failure in both settings, with a lead time of 8-10 hours, demonstrating the model’s promise in supporting early warning systems designed to anticipate respiratory compromise. A significant advantage of our approach is its compatibility with existing hospital telemetry infrastructure, suggesting that enhanced respiratory monitoring can be achieved with minimal additional costs and reduced clinical workload. Furthermore, the model’s ability to make predictions from a single lead and robust performance across different ECG leads imply that its performance may translate to ambulatory monitoring devices for use in remote patient monitoring and hospital-at-home programs.

Comparing to other RR predictors in the literature is not straightforward, primarily due to large differences in the scale of training datasets or the popular use of PPG instead of ECG. For example, Zhao et. al.\cite{zhao:2023:ecg_ppg_transformer_rr} train on combined ECG and PPG to achieve an MAE of 1.20 bpm by making use of MIMIC-II\cite{lee:2011:mimic_ii} and CapnoBase\cite{karlen:2010:capnobase} which contain 53 and 43 patients respectively, with 8 minutes of data per patient. Crucially this result employs subject-level ten-fold cross-validation instead of a completely external validation set. Another study by Roberts et. al.\cite{roberts:2024:open_source_rr} use just ECG and report a RMSE of 2.2 bpm on 13 human subjects, with an additional validation on ECG from three ischemic hearts of anesthetized, mechanically ventilated sheep, where they achieve an impressive MAE of 0.07 bpm. Finally, Baker et. al.\cite{baker:2021:sqi_rr}, again using both PPG and ECG, achieve RMSE and MAE of 1.575 and 0.638 bpm respectively. Importantly, their approach is to perform robust signal quality indexing on the input waveforms, which rejects many inputs for being excessively noisy. From 8781 initial records they keep 1300 ``good'' 60 second segments.

The present study's scale and the model's reliance on a single lead of ECG are important distinguishing feature from these prior works. The MGH training and validation datasets were comprised of more than 20000 patients and 73 million minutes of ECG telemetry. The external validation dataset alone, collected from a distinct hospital with distinct ECG monitors and impedance pneumography devices, contained 16 million 60-second segments -- orders of magnitude larger than published training sets. To maximize the generalizability and ease-of-use of the model, we used minimal filtering of the ECG input. The primary exclusion criterion during training was excessive variance of the ImP labels over a 60\,s interval, thereby focusing on label quality rather than input signal quality. Despite using ECG signals with minimal preprocessing, the model achieved performance comparable to or exceeding previous studies while offering substantially greater generalizability due to the sheer scale of the training dataset and rigorous external validation approach.

It is worth emphasizing the potential generality of the approach taken in this work. Taking a broader view, any relevant hospital vital sign, test result or diagnosis that can be aligned with ECG telemetry with sufficient accuracy is, in principle, amenable to the same analysis as performed in this paper. Of course not all information can be predicted from ECG (See \cite{alcaraz:2024:ecg_labs, alcaraz:2024:ecg_diagnosis} for bulk studies of ECG's ability to predict various lab results and clinical diagnoses), but combining this with PPG and blood pressure (BP) waveforms has the potential to predict, within some margin of error, many important clinical quantities. For some predictions this may even be enhanced by adding tabular data, such as demographics, diagnoses and details of a patient's medical history.

An aspect of this work that requires further study is bridging the gap from simple cohort differentiation to early warning metrics in individual patients. The results in this work suggest that some metric based on a smooth increase of RR over time (see patient 1 in panel B of Fig.~\ref{fig:floor_rr_relative_change_composite}, and note the monotonicity of increasing bars in
Fig.~\ref{fig:floor_rr_relative_change}
and Fig.~\ref{fig:sts_relative_change_many_ref})
or the search for notable step functions in RR (see patient 2 in panel B of Fig.~\ref{fig:floor_rr_relative_change_composite}) would be potential targets. Some patients, however, seem to exhibit few changes (see panel A in Fig.~\ref{fig:floor_rr_relative_change_composite}). It is likely that optimal early warning systems would necessitate a second level of modeling which also incorporates lab results, and demographics alongside the continuous RR offered up by this model. These points will be explored in future work.

Beyond early warning systems, this work has broad significance for understanding respiratory rate physiology in complex conditions such as heart failure, neurologic injury, sepsis and circulatory shock, as well as understanding circadian rhythm physiology in hospitalized patients. The wide availability of continuous, high resolution respiratory rate dynamics on hospitalized patients should enable many new studies at a scale previously impossible to achieve.

The RR prediction model (and other models developed by a similar methodology) has several limitations to address before it can be translated into clinical practice. Perhaps most importantly is the lack of interpretability of the neural network outputs, which currently come with no confidence intervals or other guarantees. Before deploying such a model it would be important to have other systems (maybe other neural networks) to reject excessively noisy input signals, or otherwise quantify the reliability of the output of the model. A second limitation on the clinical analysis is the purely retrospective nature of the study. While care has been taken to have a \textit{light touch} with the analysis, i.e. having very few inclusion and exclusion criteria for patients, it is still wise to exercise caution in extrapolating from the results to prospective data analysis. Especially with the vision of building next generation early warning systems, it would be important to have more prospective validation and randomized trials of any algorithms developed.

In conclusion, this work demonstrates the value of combining deep learning with typical hospital monitoring systems. Continuous signals are typically only observed sporadically by busy medical staff, but neural networks are effective in ingesting these continuous signals and providing otherwise lost insights. In this work we demonstrate that respiratory rate can be accurately annotated continuously, but this methodology can be used on many other labels \cite{kite:2024:telemetry_ecg}. With further study, this technology could pave the way to better hospital-wide early warning systems.

\section{Methods}
\subsection{Dataset Curation}
\label{subsec:dataset_curation}
The data used in this work arises from retrospective archives of continuous telemetry monitoring data, either from MGH or from the open source MIMIC III waveforms database\cite{Alistair:2016:mimic3_wfdb}. In each case the data selection and filtering criteria are the same. These two data sources importantly represent independent hospitals with distinct monitoring systems.

The MGH archive contains telemetry waveforms, vitals and electronic health record (EHR) level data for $\sim$150,000 patients with hospital stays between 2014 and 2023. However, the overlap of RR labels with ECG immediately restricts this study to a cohort of $\sim$20,000 patients, partly due to the implicit restriction to ICU patients (see Table~\ref{tab:cohort_stats}). The MIMIC archive (using the matched subset) contains $\sim$10,000 patients, reducing to $\sim$ 4,500 after aligning ECG and RR. The final number of ECG-RR minutes totals almost 100 million, with individual subset numbers reported in Fig.~\ref{fig:graphical_abstract}.

\subsubsection{Electrocardiogram}
\label{subsec:ecg_filter}
ECG is selected by looking for continuous uninterrupted minutes of data following a removal of values in excess of $\pm$60\,mV. Minutes with any flat leads are also excluded. Signals are downsampled to 120\,Hz, meaning each segment is 7200 indices long. Upon feeding an ECG array to the NN each segment is normalized by subtracting its mean and dividing by its standard deviation. Otherwise, no further filtering is applied. This minimalist approach to data curation facilitates a future application of this technology to machine learning-enhanced real-time monitoring, allowing neural networks to digest signals as they are currently recorded by standard-of-care monitoring devices, and with very few segments meeting limited exclusion criteria.

ECG from MGH always has four available leads, while data within MIMIC more often has one or two concurrent leads. To enable external validation, the neural network was trained on single lead data, and all evaluations are applied to all available leads.

\subsubsection{Respiration rate}
These minutes of ECG are aligned in time with the automated RR labels, which are recorded at 0.5\,Hz at MGH (both ImP and capnography) or 1\,Hz with MIMIC. In both cases the RR is averaged over the minute, and the minute of data is accepted if and only if the following criteria are met:
\begin{itemize}[noitemsep, topsep=0pt, parsep=0pt, partopsep=0pt]
    \item The minimum reported RR value is >0 bpm.
    \item The minute-averaged RR value is $\geq$10 bpm and $\leq$50 bpm.
    \item The maximum and minimum RR values are <10 bpm apart.
    \item The standard deviation of RR values across the minute is <2 bpm.
\end{itemize}

\subsubsection{Clinical data}
\label{subsec:clinical_data_methods}
Times of intubation are found by inspecting a human-annotated chart field \texttt{oxygen\_device}, and noting where this transitioned from a non-invasive key (e.g. \texttt{None\_(Room\_air)}, \texttt{Nasal\_cannula}, \texttt{Aerosol\_mask}, \texttt{Face\_tent}, etc.) to the key \texttt{Ventilator}. This label can have minutes of delay, and as such the timestamp is adjusted backwards to the first machine-annotated value of \texttt{vent\_rate} or \texttt{set\_tv}, both ventilator settings. These machine values are also stored in the retrospective archive, allowing for a highly precise corroboration of human-annotated values. A further 5 minute \textit{grace period} was subtracted from all event times so as to avoid the analysis being confounded by sudden changes in the ECG marking the occurrence of the event itself (i.e. a lead time of 2 hours in Fig.~\ref{fig:floor_rr_relative_change_composite} is actually 2 hours and 5 minutes).

In the reintubation cohort, respiratory failure is determined as the main cause by inspecting blood tests in the retrospective archive, which are frequently taken for post-op patients. These tests measure blood \texttt{pH}, and the partial pressures \texttt{pO2} and \texttt{pCO2}. A blood test is labeled as consistent with respiratory failure if and only if one of the following three criteria are satisfied:
\texttt{pO2} < 100\,mmHG, \texttt{pCO2} > 50\,mmHG, or \texttt{pH} < 7.3.

\subsection{Acknowledgments}
The authors thank members of the Division of Cardiac Surgery at the Massachusetts General Hospital (MGH) for providing access to outcomes data adjudicated for the Society for Thoracic Surgeons (STS) institutional database. A.D.A. acknowledges funding from the Wellman Center at Massachusetts General Hospital and from the NIH (R01HL173544).

\subsection{Author contributions}
T.K. and A.D.A. conceived and designed the study. T.K. performed data gathering and curation, designed and implemented the machine learning models and training protocol, conducted statistical analysis, and validated model performance. T.K. and B.A. led data analysis and interpretation of clinical cohorts. N.H. provided critical feedback and guidance throughout the project. T.M.S. and A.A.O. curated the STS database used within the second clinical analysis. T.K., B.A., and A.D.A. prepared the manuscript. A.D.A. and B.A. supervised the research. All authors reviewed and approved the final manuscript.

\subsection{Data availability statement}
The patient data from the derivation cohort used for this study is part of an institutional data repository and electronic health record with protected health information and cannot be uniformly released for open-source use. Data used for external validation is publicly available through the Medical Information Mart for Intensive Care (MIMIC) III database, with links provided in the references\cite{Alistair:2016:mimic3_wfdb}. More detailed data access to the derivation cohort will require institutional review board approval and relevant data use agreement from Mass General Brigham. Inquiries regarding data availability can be directed to the corresponding author.

\subsection{Additional Information}
T.K., B.A., and A.D.A. are inventors on a patent application covering material from this manuscript and filed by The General Hospital Corporation with the U.S. Patent and Trademark Office. The remaining authors declare no competing interests.

\bibliography{lit}

\begin{thebibliography}{10}
\urlstyle{rm}
\expandafter\ifx\csname url\endcsname\relax
  \def\url#1{\texttt{#1}}\fi
\expandafter\ifx\csname urlprefix\endcsname\relax\def\urlprefix{URL }\fi
\expandafter\ifx\csname doiprefix\endcsname\relax\def\doiprefix{DOI: }\fi
\providecommand{\bibinfo}[2]{#2}
\providecommand{\eprint}[2][]{\url{#2}}

\bibitem{rodwin:2020:preventable_mortality}
\bibinfo{author}{Rodwin, B.~A.} \emph{et~al.}
\newblock \bibinfo{journal}{\bibinfo{title}{Rate of preventable mortality in hospitalized patients: a systematic review and meta-analysis}}.
\newblock {\emph{\JournalTitle{Journal of General Internal Medicine}}} \textbf{\bibinfo{volume}{35}}, \bibinfo{pages}{2099--2106}, \doiprefix\url{10.1007/s11606-019-05592-5} (\bibinfo{year}{2020}).

\bibitem{makary:2016:medical_error}
\bibinfo{author}{Makary, M.~A.} \& \bibinfo{author}{Daniel, M.}
\newblock \bibinfo{journal}{\bibinfo{title}{Medical error—the third leading cause of death in the us}}.
\newblock {\emph{\JournalTitle{BMJ}}} \textbf{\bibinfo{volume}{353}}, \bibinfo{pages}{i2139}, \doiprefix\url{10.1136/bmj.i2139} (\bibinfo{year}{2016}).

\bibitem{hall:2016:burnout_healthcare}
\bibinfo{author}{Hall, L.~H.}, \bibinfo{author}{Johnson, J.}, \bibinfo{author}{Watt, I.}, \bibinfo{author}{Tsipa, A.} \& \bibinfo{author}{O'Connor, D.~B.}
\newblock \bibinfo{journal}{\bibinfo{title}{Healthcare staff wellbeing, burnout, and patient safety: A systematic review}}.
\newblock {\emph{\JournalTitle{PLOS ONE}}} \textbf{\bibinfo{volume}{11}}, \bibinfo{pages}{e0159015}, \doiprefix\url{10.1371/journal.pone.0159015} (\bibinfo{year}{2016}).

\bibitem{cretikos:2007:rr_matters}
\bibinfo{author}{Cretikos, M.} \emph{et~al.}
\newblock \bibinfo{journal}{\bibinfo{title}{The objective medical emergency team activation criteria: a case-control study}}.
\newblock {\emph{\JournalTitle{Resuscitation}}} \textbf{\bibinfo{volume}{73}}, \bibinfo{pages}{62--72}, \doiprefix\url{10.1016/j.resuscitation.2006.08.020} (\bibinfo{year}{2007}).

\bibitem{lovett:2005:vexatious_vital}
\bibinfo{author}{Lovett, P.~B.}, \bibinfo{author}{Buchwald, J.~M.}, \bibinfo{author}{St{\"u}rmann, K.} \& \bibinfo{author}{Bijur, P.}
\newblock \bibinfo{journal}{\bibinfo{title}{The vexatious vital: neither clinical measurements by nurses nor an electronic monitor provides accurate measurements of respiratory rate in triage}}.
\newblock {\emph{\JournalTitle{Annals of Emergency Medicine}}} \textbf{\bibinfo{volume}{45}}, \bibinfo{pages}{68--76}, \doiprefix\url{10.1016/j.annemergmed.2004.06.016} (\bibinfo{year}{2005}).

\bibitem{cretikos:2008:neglected_vital}
\bibinfo{author}{Cretikos, M.~A.} \emph{et~al.}
\newblock \bibinfo{journal}{\bibinfo{title}{Respiratory rate: the neglected vital sign}}.
\newblock {\emph{\JournalTitle{Medical Journal of Australia}}} \textbf{\bibinfo{volume}{188}}, \bibinfo{pages}{657--659}, \doiprefix\url{10.5694/j.1326-5377.2008.tb01825.x} (\bibinfo{year}{2008}).

\bibitem{duckitt:2007:early_warning}
\bibinfo{author}{Duckitt, R.~W.} \emph{et~al.}
\newblock \bibinfo{journal}{\bibinfo{title}{Worthing physiological scoring system: derivation and validation of a physiological early-warning system for medical admissions. an observational, population-based single-centre study}}.
\newblock {\emph{\JournalTitle{British Journal of Anaesthesia}}} \textbf{\bibinfo{volume}{98}}, \bibinfo{pages}{769--774}, \doiprefix\url{10.1093/bja/aem097} (\bibinfo{year}{2007}).

\bibitem{moss:2017:manual_rr_for_ews}
\bibinfo{author}{Moss, T.~J.} \emph{et~al.}
\newblock \bibinfo{journal}{\bibinfo{title}{Cardiorespiratory dynamics measured from continuous ecg monitoring improves detection of deterioration in acute care patients: A retrospective cohort study}}.
\newblock {\emph{\JournalTitle{PLOS ONE}}} \textbf{\bibinfo{volume}{12}}, \bibinfo{pages}{e0181448}, \doiprefix\url{10.1371/journal.pone.0181448} (\bibinfo{year}{2017}).

\bibitem{charlton:2018:breathing_review}
\bibinfo{author}{Charlton, P.~H.} \emph{et~al.}
\newblock \bibinfo{journal}{\bibinfo{title}{Breathing rate estimation from the electrocardiogram and photoplethysmogram: A review}}.
\newblock {\emph{\JournalTitle{IEEE Reviews in Biomedical Engineering}}} \textbf{\bibinfo{volume}{11}}, \bibinfo{pages}{2--20}, \doiprefix\url{10.1109/RBME.2017.2763681} (\bibinfo{year}{2018}).

\bibitem{kite:2024:telemetry_ecg}
\bibinfo{author}{{Kite}, T.}, \bibinfo{author}{{Tahamid Siam}, U.}, \bibinfo{author}{{Ayers}, B.}, \bibinfo{author}{{Houstis}, N.} \& \bibinfo{author}{{Aguirre}, A.~D.}
\newblock \bibinfo{journal}{\bibinfo{title}{{Unlocking Telemetry Potential: Self-Supervised Learning for Continuous Clinical Electrocardiogram Monitoring}}}.
\newblock {\emph{\JournalTitle{arXiv e-prints}}} \bibinfo{pages}{arXiv:2406.16915}, \doiprefix\url{10.48550/arXiv.2406.16915} (\bibinfo{year}{2024}).
\newblock \eprint{2406.16915}.

\bibitem{charlton:2017:rr_extraction}
\bibinfo{author}{Charlton, P.~H.} \emph{et~al.}
\newblock \bibinfo{journal}{\bibinfo{title}{Extraction of respiratory signals from the electrocardiogram and photoplethysmogram: technical and physiological determinants}}.
\newblock {\emph{\JournalTitle{Physiological Measurement}}} \textbf{\bibinfo{volume}{38}}, \bibinfo{pages}{669--690}, \doiprefix\url{10.1088/1361-6579/aa670e} (\bibinfo{year}{2017}).

\bibitem{Alistair:2016:mimic3_wfdb}
\bibinfo{author}{Johnson, A. E.~W.} \emph{et~al.}
\newblock \bibinfo{journal}{\bibinfo{title}{Mimic-iii, a freely accessible critical care database}}.
\newblock {\emph{\JournalTitle{Scientific Data}}} \textbf{\bibinfo{volume}{3}}, \bibinfo{pages}{160035}, \doiprefix\url{10.1038/sdata.2016.35} (\bibinfo{year}{2016}).

\bibitem{zhao:2023:ecg_ppg_transformer_rr}
\bibinfo{author}{Zhao, Q.} \emph{et~al.}
\newblock \bibinfo{journal}{\bibinfo{title}{Predicting respiratory rate from electrocardiogram and photoplethysmogram using a transformer-based model}}.
\newblock {\emph{\JournalTitle{Bioengineering (Basel, Switzerland)}}} \textbf{\bibinfo{volume}{10}}, \bibinfo{pages}{1024}, \doiprefix\url{10.3390/bioengineering10091024} (\bibinfo{year}{2023}).

\bibitem{lee:2011:mimic_ii}
\bibinfo{author}{Lee, J.} \emph{et~al.}
\newblock \bibinfo{title}{Open-access mimic-ii database for intensive care research}.
\newblock In \emph{\bibinfo{booktitle}{2011 Annual International Conference of the IEEE Engineering in Medicine and Biology Society}}, \bibinfo{pages}{8315--8318}, \doiprefix\url{10.1109/IEMBS.2011.6092050} (\bibinfo{year}{2011}).

\bibitem{karlen:2010:capnobase}
\bibinfo{author}{Karlen, W.}, \bibinfo{author}{Turner, M.}, \bibinfo{author}{Cooke, E.}, \bibinfo{author}{Dumont, G.} \& \bibinfo{author}{Ansermino, J.~M.}
\newblock \bibinfo{title}{Capnobase: Signal database and tools to collect, share and annotate respiratory signals}.
\newblock In \emph{\bibinfo{booktitle}{Proceedings of the Annual Meeting of the Society for Technology in Anesthesia (STA)}} (\bibinfo{year}{2010}).

\bibitem{roberts:2024:open_source_rr}
\bibinfo{author}{Roberts, J.~D.}, \bibinfo{author}{Walton, R.~D.}, \bibinfo{author}{Loyer, V.} \emph{et~al.}
\newblock \bibinfo{journal}{\bibinfo{title}{Open-source software for respiratory rate estimation using single-lead electrocardiograms}}.
\newblock {\emph{\JournalTitle{Scientific Reports}}} \textbf{\bibinfo{volume}{14}}, \bibinfo{pages}{167}, \doiprefix\url{10.1038/s41598-023-50470-0} (\bibinfo{year}{2024}).

\bibitem{baker:2021:sqi_rr}
\bibinfo{author}{Baker, S.}, \bibinfo{author}{Xiang, W.} \& \bibinfo{author}{Atkinson, I.}
\newblock \bibinfo{journal}{\bibinfo{title}{Determining respiratory rate from photoplethysmogram and electrocardiogram signals using respiratory quality indices and neural networks}}.
\newblock {\emph{\JournalTitle{PLOS ONE}}} \textbf{\bibinfo{volume}{16}}, \bibinfo{pages}{e0249843}, \doiprefix\url{10.1371/journal.pone.0249843} (\bibinfo{year}{2021}).

\bibitem{alcaraz:2024:ecg_labs}
\bibinfo{author}{{Lopez Alcaraz}, J.~M.} \& \bibinfo{author}{{Strodthoff}, N.}
\newblock \bibinfo{journal}{\bibinfo{title}{{CardioLab: Laboratory Values Estimation from Electrocardiogram Features -- An Exploratory Study}}}.
\newblock {\emph{\JournalTitle{arXiv e-prints}}} \bibinfo{pages}{arXiv:2407.18629}, \doiprefix\url{10.48550/arXiv.2407.18629} (\bibinfo{year}{2024}).
\newblock \eprint{2407.18629}.

\bibitem{alcaraz:2024:ecg_diagnosis}
\bibinfo{author}{{Lopez Alcaraz}, J.~M.} \& \bibinfo{author}{{Strodthoff}, N.}
\newblock \bibinfo{journal}{\bibinfo{title}{{Estimation of Cardiac and Non-cardiac Diagnosis from Electrocardiogram Features}}}.
\newblock {\emph{\JournalTitle{arXiv e-prints}}} \bibinfo{pages}{arXiv:2408.17329}, \doiprefix\url{10.48550/arXiv.2408.17329} (\bibinfo{year}{2024}).
\newblock \eprint{2408.17329}.

\bibitem{Liu:2022:convnext_original}
\bibinfo{author}{{Liu}, Z.} \emph{et~al.}
\newblock \bibinfo{journal}{\bibinfo{title}{{A ConvNet for the 2020s}}}.
\newblock {\emph{\JournalTitle{arXiv e-prints}}} \bibinfo{pages}{arXiv:2201.03545}, \doiprefix\url{10.48550/arXiv.2201.03545} (\bibinfo{year}{2022}).
\newblock \eprint{2201.03545}.

\bibitem{Zachi:2019:ecg_resnet_afib_original}
\bibinfo{author}{Attia, Z.~I.} \emph{et~al.}
\newblock \bibinfo{journal}{\bibinfo{title}{An artificial intelligence-enabled ecg algorithm for the identification of patients with atrial fibrillation during sinus rhythm: a retrospective analysis of outcome prediction}}.
\newblock {\emph{\JournalTitle{The Lancet}}} \textbf{\bibinfo{volume}{394}}, \bibinfo{pages}{861--867}, \doiprefix\url{10.1016/S0140-6736(19)31721-0} (\bibinfo{year}{2019}).

\bibitem{Zachi:2019:ecg_age_sex}
\bibinfo{author}{Attia, Z.~I.} \emph{et~al.}
\newblock \bibinfo{journal}{\bibinfo{title}{Age and sex estimation using artificial intelligence from standard 12-lead ecgs}}.
\newblock {\emph{\JournalTitle{Circ Arrhythm Electrophysiol}}} \textbf{\bibinfo{volume}{12}}, \bibinfo{pages}{e007284}, \doiprefix\url{10.1161/CIRCEP.119.007284} (\bibinfo{year}{2019}).
\newblock \bibinfo{note}{Epub 2019 Aug 27}.

\bibitem{Ribeiro:2020:ecg_resnet}
\bibinfo{author}{Ribeiro, A.~H.} \emph{et~al.}
\newblock \bibinfo{journal}{\bibinfo{title}{Automatic diagnosis of the 12-lead ecg using a deep neural network}}.
\newblock {\emph{\JournalTitle{Nature Communications}}} \textbf{\bibinfo{volume}{11}}, \bibinfo{pages}{1760}, \doiprefix\url{10.1038/s41467-020-15432-4} (\bibinfo{year}{2020}).

\bibitem{Kiyasseh:2020:clocs_original}
\bibinfo{author}{{Kiyasseh}, D.}, \bibinfo{author}{{Zhu}, T.} \& \bibinfo{author}{{Clifton}, D.~A.}
\newblock \bibinfo{journal}{\bibinfo{title}{{CLOCS: Contrastive Learning of Cardiac Signals Across Space, Time, and Patients}}}.
\newblock {\emph{\JournalTitle{arXiv e-prints}}} \bibinfo{pages}{arXiv:2005.13249}, \doiprefix\url{10.48550/arXiv.2005.13249} (\bibinfo{year}{2020}).
\newblock \eprint{2005.13249}.

\bibitem{Abbaspourazad:2023:foundation_wearable}
\bibinfo{author}{{Abbaspourazad}, S.} \emph{et~al.}
\newblock \bibinfo{journal}{\bibinfo{title}{{Large-scale Training of Foundation Models for Wearable Biosignals}}}.
\newblock {\emph{\JournalTitle{arXiv e-prints}}} \bibinfo{pages}{arXiv:2312.05409}, \doiprefix\url{10.48550/arXiv.2312.05409} (\bibinfo{year}{2023}).
\newblock \eprint{2312.05409}.

\bibitem{Vaswani:2017:attention_original}
\bibinfo{author}{{Vaswani}, A.} \emph{et~al.}
\newblock \bibinfo{journal}{\bibinfo{title}{{Attention Is All You Need}}}.
\newblock {\emph{\JournalTitle{arXiv e-prints}}} \bibinfo{pages}{arXiv:1706.03762}, \doiprefix\url{10.48550/arXiv.1706.03762} (\bibinfo{year}{2017}).
\newblock \eprint{1706.03762}.

\bibitem{Chollet:2016:xception_depthwise_separable}
\bibinfo{author}{{Chollet}, F.}
\newblock \bibinfo{journal}{\bibinfo{title}{{Xception: Deep Learning with Depthwise Separable Convolutions}}}.
\newblock {\emph{\JournalTitle{arXiv e-prints}}} \bibinfo{pages}{arXiv:1610.02357}, \doiprefix\url{10.48550/arXiv.1610.02357} (\bibinfo{year}{2016}).
\newblock \eprint{1610.02357}.

\bibitem{Hendrycks:2016:gelu_original}
\bibinfo{author}{{Hendrycks}, D.} \& \bibinfo{author}{{Gimpel}, K.}
\newblock \bibinfo{journal}{\bibinfo{title}{{Gaussian Error Linear Units (GELUs)}}}.
\newblock {\emph{\JournalTitle{arXiv e-prints}}} \bibinfo{pages}{arXiv:1606.08415}, \doiprefix\url{10.48550/arXiv.1606.08415} (\bibinfo{year}{2016}).
\newblock \eprint{1606.08415}.

\bibitem{Ulyanov:2016:instance_norm_original}
\bibinfo{author}{{Ulyanov}, D.}, \bibinfo{author}{{Vedaldi}, A.} \& \bibinfo{author}{{Lempitsky}, V.}
\newblock \bibinfo{journal}{\bibinfo{title}{{Instance Normalization: The Missing Ingredient for Fast Stylization}}}.
\newblock {\emph{\JournalTitle{arXiv e-prints}}} \bibinfo{pages}{arXiv:1607.08022}, \doiprefix\url{10.48550/arXiv.1607.08022} (\bibinfo{year}{2016}).
\newblock \eprint{1607.08022}.

\bibitem{srivastava:2014:dropout}
\bibinfo{author}{Srivastava, N.}, \bibinfo{author}{Hinton, G.}, \bibinfo{author}{Krizhevsky, A.}, \bibinfo{author}{Sutskever, I.} \& \bibinfo{author}{Salakhutdinov, R.}
\newblock \bibinfo{journal}{\bibinfo{title}{Dropout: A simple way to prevent neural networks from overfitting}}.
\newblock {\emph{\JournalTitle{Journal of Machine Learning Research}}} \textbf{\bibinfo{volume}{15}}, \bibinfo{pages}{1929--1958} (\bibinfo{year}{2014}).

\bibitem{micikevicius:2017:mixed}
\bibinfo{author}{Micikevicius, P.} \emph{et~al.}
\newblock \bibinfo{journal}{\bibinfo{title}{Mixed precision training}}.
\newblock {\emph{\JournalTitle{arXiv preprint arXiv:1710.03740}}}  (\bibinfo{year}{2017}).

\end{thebibliography}

\appendix

\newpage
\begin{figure}[b!]
\centering
\includegraphics[width=0.8\textwidth]{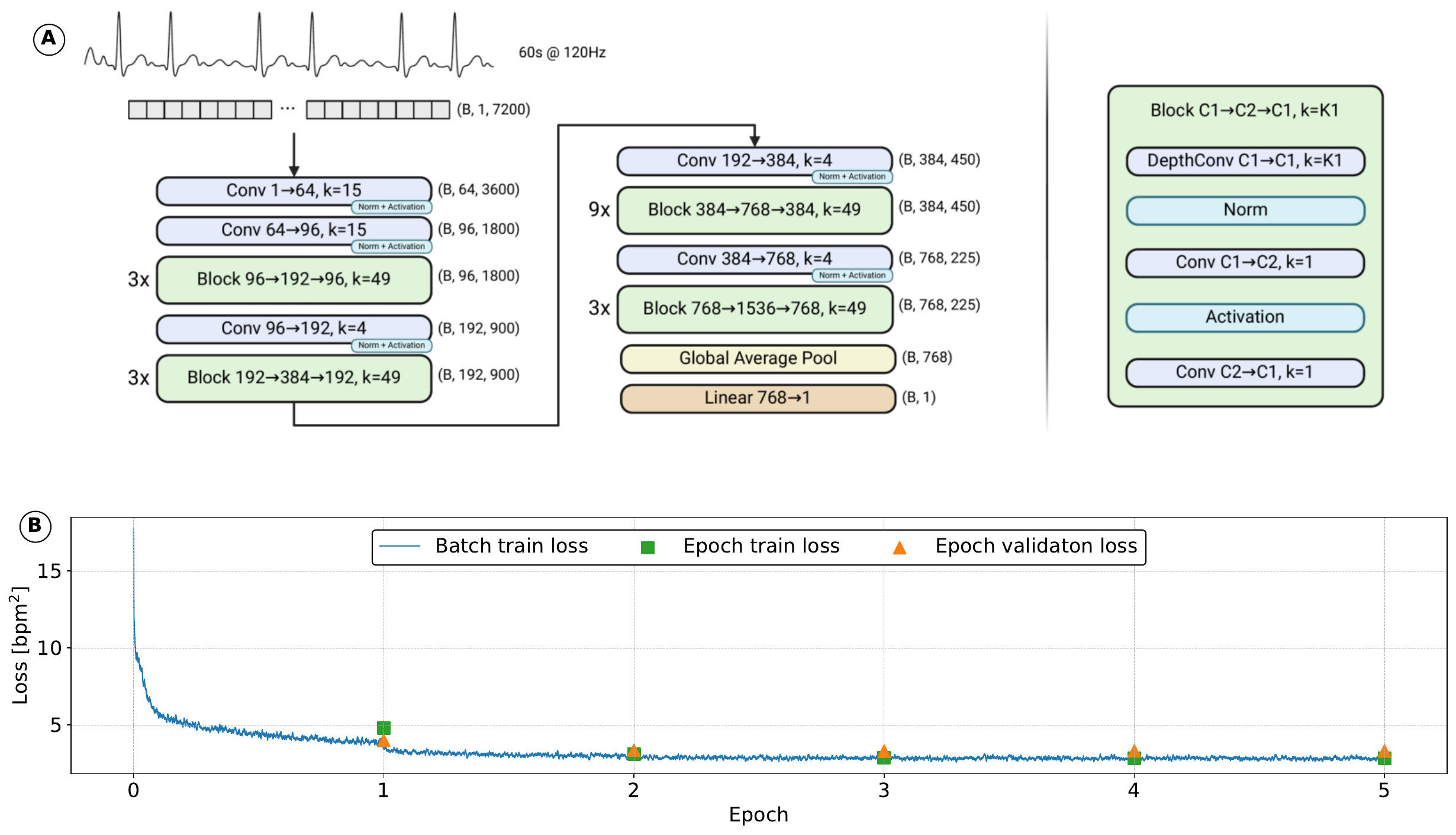}
\caption{The architecture of the neural network is composed of 60 layers and is based on the ConvNeXt architecture \cite{Liu:2022:convnext_original}. The input is a single 60\,s ECG lead sampled at 120\,Hz, and is normalized by subtracting its mean and then dividing by its standard deviation. The output is a single number prediction of the average respiration rate.}
\label{fig:architecture_and_loss}
\end{figure}
\begin{figure}
\centering
\includegraphics[width=1.0\textwidth]{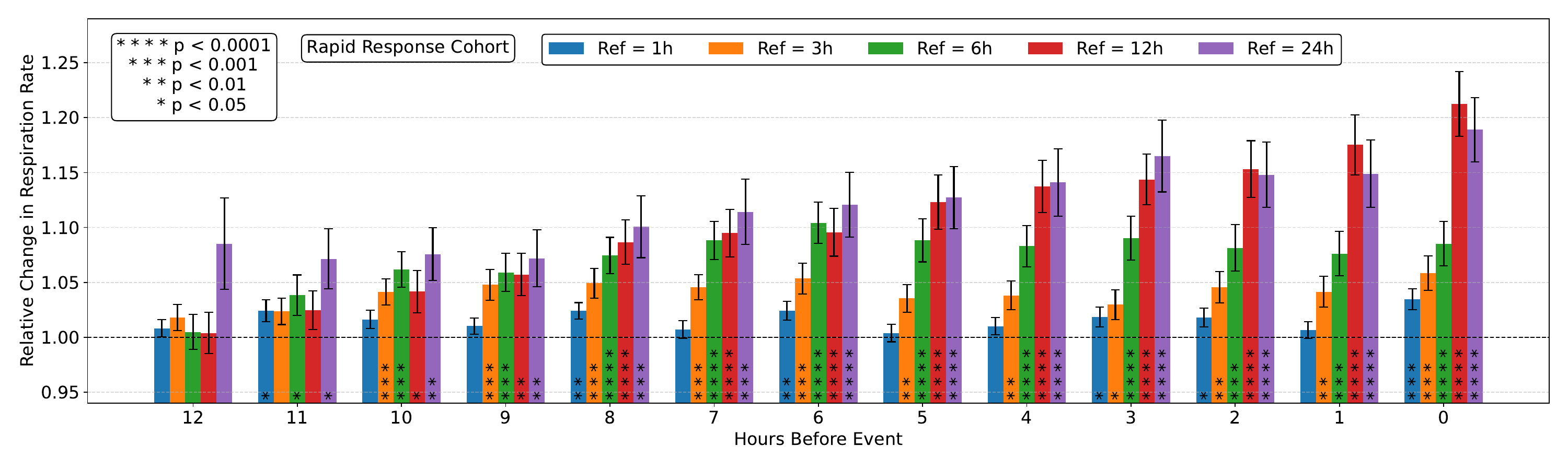}
\caption{Change in RR leading up to a rapid response. Data are displayed as in panel C from
Fig.~\ref{fig:floor_rr_relative_change_composite},
now using a range of reference times to establish the patient-specific baseline.}
\label{fig:floor_rr_relative_change}
\end{figure}
\begin{figure}
\centering
\includegraphics[width=1.0\textwidth]{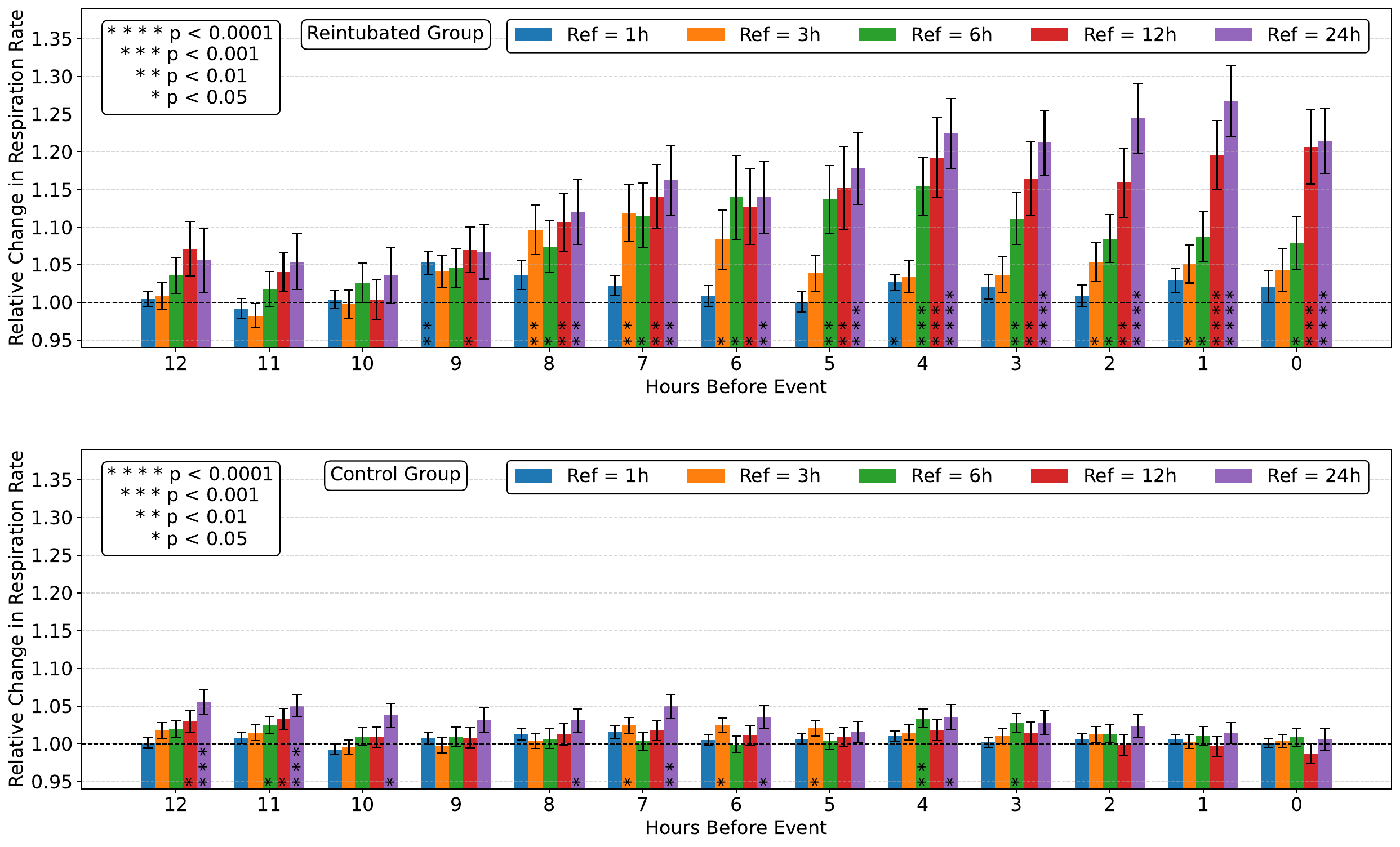}
\caption{Change in RR leading up to a postoperative reintubation. Data are displayed as in panel B from
Fig.~\ref{fig:sts_relative_change_many_ref},
now using a range of reference times to establish the patient-specific baseline.}
\label{fig:sts_relative_change_many_ref}
\end{figure}
\vspace{-1cm}
\section{Neural Network Architecture}
\label{app:nn_architecture}
We opt to analyze the ECG as a \textit{1D image}, and employ techniques from the computer vision literature. This choice more closely resembles most of the ECG-ML literature (see for example \cite{Zachi:2019:ecg_resnet_afib_original, Zachi:2019:ecg_age_sex, Ribeiro:2020:ecg_resnet, Kiyasseh:2020:clocs_original, Abbaspourazad:2023:foundation_wearable}), and is suitable given a fixed window size as ours. It is noteworthy however that the window size in this work is 60 seconds, much longer than the common 10 seconds as for the standard 10s-12 lead ECGs\cite{kite:2024:telemetry_ecg}. It is therefore important to have a much wider receptive field to compare many ECG peaks in deriving RR, while not excessively inflating the parameter count. Given these considerations, the chosen neural network architecture is based on on ConvNeXt \cite{Liu:2022:convnext_original}, which has been shown to be competitive with Transformers \cite{Vaswani:2017:attention_original} in computer vision tasks, and was found to be much more efficient for our 1D task than other attention-based methods. One key component of the ConvNeXt architecture is the use of depthwise separable convolutions \cite{Chollet:2016:xception_depthwise_separable}, which are very parameter efficient and allow for large kernel sizes without incurring a severe parameter cost, especially if combined with dilation.

The neural network architecture is depicted in panel A of Fig.~\ref{fig:architecture_and_loss}, with tensor shapes at each stage shown. All activation layers employ Gaussian Error Linear Unit (GELU) \cite{Hendrycks:2016:gelu_original}, and normalization layers employ InstanceNorm \cite{Ulyanov:2016:instance_norm_original}. All convolution layers (blue) are padded and followed by strided average pools for downsampling where needed. All convolutions outside of ConvNeXt blocks (green) are followed by a normalization and activation layer. Within the blocks we use the ConvNeXt convention, which reduces the number of normalizations and activations to more closely match transformers \cite{Liu:2022:convnext_original}. Importantly the depthwise convolution (DepthConv) doesn't allow channels to communicate with each other. This means that, heuristically speaking, the ConvNeXt block first gathers local information across time in the ECG in each channel, then performs some calculation across all channels but at each individual pixel by expanding and shrinking the channel dimension across two layers with an activation layer (essentially an MLP applied in each pixel across all channels, much like a transformer's feed forward layer \cite{Vaswani:2017:attention_original}). The final network layers are a global average pool, which aggregates each channel across all of the ECG length, and a linear layer which maps these channels to a single output. Dropout layers\cite{srivastava:2014:dropout} with $p=0.3$ are applied after each convolution. The final choice of architecture contains, 60 layers and 14.91 million trainable parameters.

We train the model for five epochs, where each epoch loads every minute of data but selects a single random lead of the four available. It is often important when working with biosignals to split epochs at patient level, not segment level. However, the quantity of data available here combined with the dynamic nature of respiratory rate showed this was not necessary (note that train, validation and test data splits are indeed separated at patient level, which \textit{is} essential for reliable results). The model trains by backpropagating the mean square error (MSE) of the current batch, and updating model parameters. The batch size is 128 ECG segments. For every epoch the learning rate is reduced by a factor of 10, starting from $10^{-3}$. A small weight decay of $5\times 10^{-5}$ is applied. Mixed precision training\cite{micikevicius:2017:mixed} is employed to allow larger batch sizes on the same device, a single RTX 4090. The training curve in panel B of Fig.~\ref{fig:architecture_and_loss} shows stable training, with a very small amount of final overfitting and almost optimal results within two epochs. This suggests that the task is in a very data-rich regime, however a larger model was not able to achieve better results. This hints that performance is close to the optimal predictor for this task, especially given the MAE on an external validation set of $<2$ bpm, which is likely approaching the accuracy of the labels from ImP.

\section{Clinical cohorts further analysis}
\label{app:more_ref_hours}
In the main text's Figs~4 and 5
a reference point of 12 hours was used to normalize each patient's RR. This design choice is explored in Fig.~\ref{fig:floor_rr_relative_change} for the rapid response cohort, and Fig.~\ref{fig:sts_relative_change_many_ref} for the post-op cohort, showing that most choices of normalization reveal the same statistical significance (in absence of single hour changes, which are not substantial except in the case of one hour before rapid response events). Most notably the increase appears to be monotonic building up to the event, revealing that the increase in breathing rate is a gradual rise spanning many hours. This is promising for the hope of detecting these events early.

\end {document}